\documentclass[twocolumn,amssymb,amsmath,preprintnumbers,floatfix,aps]{revtex4}
\usepackage{graphicx,axodraw,bm,psfrag}

\newcommand{\be}{\begin{equation}}
\newcommand{\ee}{\end{equation}}
\newcommand{\bear}{\begin{eqnarray}}
\newcommand{\eear}{\end{eqnarray}}
\newcommand{\ba}{\begin{array}}
\newcommand{\ea}{\end{array}}
\begin{document}

\preprint{\small\scriptsize FERMILAB-PUB-10-167-T}
\title{\boldmath CP violation in $B_s$ mixing from heavy Higgs exchange}
\author{Bogdan A. Dobrescu, Patrick J. Fox and Adam Martin} 
\affiliation{Theoretical Physics Department,
Fermi National Accelerator Laboratory, Batavia, Illinois, USA}

\date{May 24, 2010; revised June 14, 2010}

\begin{abstract}
The anomalous dimuon charge asymmetry reported by the D0 Collaboration
may be due to the tree-level exchange of some spin-0 particles that mediate CP violation in 
$B_s\!-\!\bar{B}_s$ meson mixing.  We show that for a range of couplings and masses, the heavy neutral states in 
a two Higgs doublet model can generate a large charge asymmetry. This range is natural in ``uplifted 
supersymmetry'',  and may enhance the $B^-\!\to\! \tau\nu$ and $B_s\!\to\! \mu^+\mu^-$ decay rates.  
However, we point out that on general grounds 
the reported central value of the charge asymmetry requires new physics not only in $B_s\!-\!\bar{B}_s$ mixing
but also in $\Delta B =1 $ transitions or in $B_d\!-\!\bar{B}_d$ mixing.
\end{abstract}

\maketitle

{\bf Introduction.}---The Standard Model (SM) predicts that the violation of CP symmetry in $B\! - \!\bar{B}$ meson 
mixing is very small \cite{Amsler:2008zzb}, and various  measurements have so far confirmed this prediction in the $B_d$ system. 
Experimental sensitivity to the properties of  $B_s$ mesons has improved
within the last few years, with well-understood data sets from $p\bar{p}$ collisions at the Tevatron 
analyzed by the D0 and CDF Collaborations.
The large ratio of the $s$ and $d$ quark masses, and also the large $V_{ts}/V_{td}$ ratio make the $B_s$ system more sensitive to
new physics than the $B_d$ system. 
We explore here the possibility that tree-level exchange of new particles induces a sizable
CP violation in $B_s\!-\!\bar{B}_s$ mixing. 

Recently \cite{Abazov:2010hv}, the D0 Collaboration has reported evidence for CP violation in final states involving 
two muons of the same charge, arising from semileptonic decays of $b$ hadrons.
The like-sign dimuon charge asymmetry, measured by D0 with 6.1 fb$^{-1}$ of data, is defined by 
\be
A_{\rm sl}^b \equiv \frac{N_b^{++}-N_b^{--}}{N_b^{++}+N_b^{--}}   ~~,
\ee
where $N_b^{++}$ is the number of events with two $b$ hadrons decaying into
$\mu^+ X$. The D0 result,  $A_{\rm sl}^b = - [9.57 \pm 2.51({\rm stat.}) \pm 1.46({\rm syst.})]\times 10^{-3}$ 
is 3.2$\sigma$ away from the SM prediction of $-0.2 \times 10^{-3}$.
The CDF \cite{cdf-dimuon} measurement of $A_{\rm sl}^b$, using 1.6 fb$^{-1}$ of data, has a positive central value,
$A_{\rm sl}^b = (8.0 \pm 9.0 \pm 6.8) \times 10^{-3} $, but is compatible with the D0 measurement at the 1.5$\sigma$ level
because its uncertainties are 4 times larger than those of D0. Combining in quadrature (including the systematic errors) 
the D0 and CDF results for $A_{\rm sl}^b$ we find a 3$\sigma$ deviation from the SM:
\be
A_{\rm sl}^b \simeq - (8.5 \pm 2.8)\times 10^{-3}  ~~.
\label{Aslb}
\ee

Another test of CP violation in $B_s\! -\!\bar{B}_s$ mixing is provided by the measurement
of the ``wrong-charge" asymmetry in semileptonic $B_s$ decays,
\be
a_{\rm sl}^s \equiv \frac{\Gamma(\bar{B}_s \to \mu^+ X)-\Gamma(B_s \to \mu^-X)}{\Gamma(\bar{B}_s \to \mu^+ X)+\Gamma(B_s \to \mu^-X)} ~~.
\ee
The D0 measurement in this channel  \cite{Abazov:2009wg}, $a_{\rm sl}^s = - (1.7 \pm 9.1^{+1.4}_{-1.5}) \times 10^{-3}$, 
is consistent with the SM.
Assuming that the CP asymmetry in $B_d\!-\!\bar{B}_d$ mixing
is negligible, the like-sign dimuon charge asymmetry is entirely due to $B_s\!-\!\bar{B}_s$ mixing and 
is related to $a_{\rm sl}^s$:
$A_{\rm sl}^b = (0.494\pm 0.043) \,  a_{\rm sl}^s $,
where the coefficient depends on the fraction of $\bar{b}$ anti-quarks which hadronize into a 
$B_s$ meson \cite{Abazov:2010hv}.
This allows the extraction of $a_{\rm sl}^s$ from Eq.~(\ref{Aslb}), which then can be combined with 
the D0 measurement of $a_{\rm sl}^s$, resulting in
\be
(a_{\rm sl}^s)_{\rm combined} \approx - (12.7\pm 5.0) \times 10^{-3} ~.
\label{combination}
\ee
Even though the inclusion of the CDF dimuon asymmetry and the D0 semileptonic wrong-charge asymmetry
reduces the deviation in $a_{\rm sl}^s$ derived from the D0 dimuon asymmetry, the above result
is still about 2.5$\sigma$ away from the SM value \cite{Lenz:2006hd}
of $(a_{\rm sl}^s)_{\rm SM} \approx 0.02 \times 10^{-3}$.

The D0 \cite{:2008fj} and CDF \cite{CDF-fpcp} Collaborations have also reconstructed $B_s\!\!\to\!\! J/\psi\, \phi$ decays and 
measured angular distributions as a function of decay time, and reported some deviation consistent with CP violation in 
$B_s\!-\!\bar{B}_s$ oscillations (see \cite{Bona:2008jn} for a fit to earlier $B_s$ data). 
The sign \cite{Ligeti:2006pm}
  and size of this deviation are compatible with Eq.~(\ref{combination}),
further strengthening the case for physics beyond the SM. 

\bigskip
{\bf Generic new physics.}---The matrix element of some new physics Hamiltonian, ${\cal H}^{\rm NP}$, 
contributing to $B_s\!-\!\bar{B}_s$ mixing may be 
parameterized as \cite{Grossman:2006ce},\cite{Lenz:2006hd},\cite{Bona:2008jn}
\be
\langle \overline{B}_s |{\cal H}^{\rm NP}  |  B_s \rangle = \left( C_{B_s} e^{-i\phi_s} - 1 \right) 2 M_{B_s} \, (M_{12}^{\rm SM})^*  ~~,
\label{param}
\ee
where $C_{B_s} > 0 $ and $-\pi\le \phi_s \le \pi$.
The magnitude of the off-diagonal element of the $B_s-\bar{B_s}$ mass matrix due to SM box diagrams is:
$|M_{12}^{\rm SM}| \simeq (9.0 \pm 1.4) \; {\rm ps}^{-1}$, where we used the same inputs as 
in \cite{Lenz:2006hd} except for the updated values of the $B_s$ decay constant
$f_{B_s} = (231\pm 15)$ MeV and bag parameter  $B = 0.86 \pm 0.04$ 
computed on the lattice with 2+1 flavors \cite{Gamiz:2009ku}.
The phase of $M_{12}^{\rm SM}$ is negligible.

The measured mass difference of the $B_s$ mass eigenstates depends linearly on $C_{B_s}$,
$\Delta M_s = 2 |M_{12}^{\rm SM}| \, C_{B_s}$. 
The combination  \cite{Barberio:2008fa} of the CDF and D0 measurements 
is $ \Delta M_s = (17.78\pm 0.12)  \; {\rm ps}^{-1}$, so that we find
\be
C_{B_s} = 0.98 \pm 0.15 ~~.
\label{cbs}
\ee 

The semileptonic wrong-charge asymmetry is given by 
\be
a_{\rm sl}^s = \frac{ 2|\Gamma_{12}| }{\Delta M_s }\sin\phi_s ~~,
\label{relation}
\ee
where $\Gamma_{12}$ is the off-diagonal element of the $B_s\!-\!\bar{B_s}$ decay-width matrix.
New physics contributing to $\Delta B = 1$ processes 
may affect $\Gamma_{12}$, but the effects are typically negligible compared to the SM $b\! \to\! c\bar{c}s$ transition
due to tree-level $W$ exchange, which is suppressed only by $V_{cb}$.
The SM prediction for $|\Gamma_{12}|$ is given by $|\Gamma_{12}^{\rm SM}| = (1/2)(0.090\pm 0.024) \; {\rm ps}^{-1}$, 
where we again used the results of \cite{Lenz:2006hd} with updated values for $f_{B_s}$ and $B$
(this is consistent with the result of \cite{Badin:2007bv}).
Using the $a_{\rm sl}^s$ value from Eq.~(\ref{combination}), we find that Eq.~(\ref{relation}) gives
\be
\sin\phi_s = -2.5 \pm 1.3 ~~.
\ee
This is a somewhat troubling result: the central value is more than 1$\sigma$ away from the physical region 
$|\sin\phi_s| \le 1$. This tension arises because the absolute value of $B_s\!-\!\bar{B}_s$ mixing is constrained 
by the measured $\Delta M_s$, not allowing enough room for an asymmetry as large as the central value 
of $a_{\rm sl}^s$ shown in Eq.~(\ref{combination}). 
This suggests that the central value of $a_{\rm sl}^s$ will be reduced by a factor of more than two 
when the error bars will become small enough. 

Alternatively, the assumptions about new physics employed here may need to be relaxed.
For example, the wrong-charge asymmetry in semileptonic $B_d$ decays, $a_{\rm sl}^d$, 
may be non-negligible. Its value given by measurements at $B$ factories is 
$(-4.7 \pm 4.6) \times 10^{-3}$ \cite{Barberio:2008fa}, so including it would change
the relation between $A_{\rm sl}^b$ and $a_{\rm sl}^s$ as
discussed in \cite{Abazov:2010hv}. This possibility is intriguing, but one should keep in mind that new physics contributions to 
$(\bar{b} d)(\bar{b} d)$ operators are often suppressed by additional powers of $m_d/m_s$ and $V_{td}/V_{ts}$
compared to those to $(\bar{b} s)(\bar{b} s)$  \cite{new}.

Another possibility is that there are sizable new contributions to $\Gamma_{12}$. 
This is problematic because the SM tree-level contribution is CKM-favored,
while new particles that induce $\Delta B =1$ effects are constrained by various limits on 
flavor-changing neutral currents ({\it e.g.}, $b\to s \gamma$ or $K\!-\!\bar{K}$ mixing) 
and by collider searches. Nevertheless, examples of relatively large shifts in $\Gamma_{12}$ can be found \cite{Badin:2007bv,Dighe:2007gt}.
Consider for example two operators, $(\bar{b}_R\gamma^\mu c_R)(\bar{u}_R\gamma^\mu s_R)$ and
$(\bar{b}_R\gamma^\mu u_R)(\bar{c}_R\gamma^\mu s_R)$, which may be induced by $W^\prime$ exchanges.
The main effect of these $\Delta B = 1$ operators
is to enhance the rate for $B_d\!\to\! DK$ decays. Given that these dominant decay modes of $B_d$ 
involve a form factor
which is not known precisely, these operators may account for a significant fraction of the measured decay width.
If the scale of the new operators is 0.9 TeV then 
 $\Gamma_{12}$ is enhanced by $30\%$.
In what follows we will focus on $\Delta B = 2$ transitions [see Eq.~(\ref{param})], ignoring
new contributions to $|\Gamma_{12}|$.

\smallskip\medskip
{\bf New physics models for $B_s\!-\!\bar{B}_s$ mixing.}---Although more experimental studies are required before concluding that 
physics beyond the SM contributes to $B_s-\bar{B}_s$ mixing, it is useful to analyze what kind of
new physics could induce CP-violating effects as large as $\sin\phi_s \approx -1$.   Given that the SM 
$B_s\!-\!\bar{B}_s$ mixing is a 1-loop effect, it is often assumed that 
new physics contributes also at one loop, for example via gluino-squark box diagrams in the 
MSSM \cite{Randall:1998te}.
However, the large effect indicated by the  data is more likely to be due to tree-level 
exchange of new particles which induce $\bar{b}s\bar{b}s$ operators.
These particles must be bosons (with spin 0, 1, or 2 being the more likely possibilities) 
carrying baryon number 0 or $\pm 2/3$. In the first case they must be 
electrically neutral and color singlets or octets. The bosons of baryon number $\pm 2/3$ 
are diquarks of electric charge $\mp 2/3$ and transform under $SU(3)_c$ 
as $\bar{3}$ or 6 (3 or $\bar{6}$ for charge +2/3).

The new bosons may be related to electroweak symmetry 
breaking, as in the case of the heavy Higgs states in two Higgs doublet models. 
We concentrate in what follows on a spin-0 boson 
$H_d^0 = (H^0 + i A^0)/\sqrt{2}$, which is electrically neutral and a color singlet (and part of a weak doublet).
The Yukawa couplings of $H_d^0$ to $b$ and $s$ quarks in the mass eigenstate basis 
are given by
\be
- H_d^0\left( y_{bs} \bar{b}_R s_L + y_{sb} \bar{s}_R b_L \right) + {\rm H.c.}
\label{interactions}
\ee
Let us assume for simplicity that the vacuum expectation value (VEV) of $H^0$ is negligible at 
tree level (the coupling to quarks induces a small VEV at one loop),
so that $H^0$ and $A^0$ have the same mass $M_A$. Examples of theories with these features are
the MSSM in the uplifted region \cite{Dobrescu:2010mk}, as discussed later, and 
composite Higgs models \cite{Chivukula:1998wd}.

Tree-level $H_d^0$ exchange gives rise to a single term in the Lagrangian  which contributes to  $B_s\!-\!\bar{B}_s$ mixing:
\be
\frac{y_{bs} y_{sb}^*}{M_A^2} (\bar{b}_R s_L)(\bar{b}_L s_R)  ~~,
\label{operator}
\ee
where the quark fields are taken in the mass eigenstate basis.
If the VEV of $H^0$ is taken into account, then additional operators contribute
\cite{Gorbahn:2009pp}, most importantly $(\bar{b}_R s_L)^2$; we will ignore these contributions  in what follows.
The matrix element of operator (\ref{operator}) is 
\be
\langle \overline{B}_s  |{\cal H}^{\rm NP}  | B_s  \rangle =  - \frac{y_{bs} y_{sb}^* \eta}{M_A^2} \, \frac{M_{B_s}^4 f^2_{B_s} B_4 }{2(m_b+m_s)^2} ~~.
\label{matrix}
\ee
The  bag parameter for operator (\ref{operator}) has been estimated  using the quenched aproximation 
on the lattice \cite{Becirevic:2001xt}, $B_4 \approx 1.16 $. The parameter $\eta \approx  4$ takes into account the running 
of  operator (\ref{operator}) between the $M_A$ and $M_{B_s}$ scales \cite{Buras:2001ra}.
For the sum of quark masses we use $m_b+m_s \approx 4.3$ GeV.  Comparing Eqs.~(\ref{param}) and (\ref{matrix}) we find
\bear
\frac{M_A}{\sqrt{|y_{bs} y_{sb}|\eta}} &=& \frac{(147 \pm 15) \; {\rm TeV}}{\left( C_{B_s}^2 + 1 - 2 C_{B_s} \cos\phi_s \right)^{1/4}} ~~,
 \nonumber \\ [1mm]
{\rm arg} (y_{bs} y_{sb}^*) &=&  \tan^{-1} \left( \frac{ C_{B_s}\sin\phi_s}{1 - C_{B_s}\cos\phi_s }\right)  ~~.  \;\;\;\;\;\;\;\;\;
\label{ma-arg}
\eear
The off-diagonal coupling $y_{bs}$ is expected to be suppressed by $V_{ts}$ compared to the diagonal $y_{b}$ Yukawa coupling
of $H_d^0$ to $\bar{b}_Rb_L$, 
while  $y_{sb}$ is suppressed by an additional factor of $m_s/m_b$, so that 
we take $|y_{bs}| \lesssim 10^{-2}$ and  $|y_{sb}| \lesssim 2\times 10^{-4}$.
When $y_{bs}$ and $y_{sb}$ saturate these upper bounds, the experimental constraint Eq.~(\ref{cbs})
on $C_{B_s}$ gives $M_A \approx  (0.65\pm 0.07) \; {\rm TeV}$  and 
${\rm arg} (y_{bs} y_{sb}^*)= -1.3 \pm 0.3$ for  
$\phi_s = -\pi/6$. Figure 1 shows the range of $M_A/\sqrt{|y_{bs} y_{sb}|}$ as a function of $\phi_s$.

\begin{figure}[t!]
\psfrag{MATeV}{$\widetilde{M}_A$ [TeV]}
\psfrag{phi}{$\phi_s$}
\psfrag{MAequalMAtimesstuff}{\hspace*{-2mm} $\widetilde{M}_A\equiv M_A \left(\frac{10^{-5}}{|y_{bs}y_{sb}|\eta}\right)^{\! 1/2}$}
\hspace*{-3mm}\includegraphics[width=0.43\textwidth]{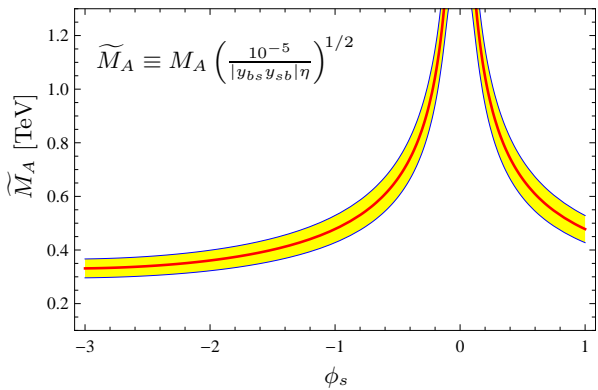}\vspace*{-3mm}
\caption{Range for $M_A$ compatible with a CP asymmetry in $B_s\!-\!\bar{B}_s$ mixing described by the $\phi_s$ angle.
The vertical size of the shaded band accounts for the 1$\sigma$ experimental uncertainty in $\Delta M_s$ and for the theoretical uncertainties  in $f_{B_s}$ and  $|M_{12}^{\rm SM}|$.   
The off-diagonal Yukawa couplings are expected to satisfy  $|y_{bs}|\lesssim 10^{-2}$ and $|y_{sb}|\lesssim 2\times 10^{-4}$.
The running between $M_A$ and $M_{B_s}$ is parametrized by $\eta \approx 4$.
 \\ [-7mm] }
\label{fig:MApsi}
\end{figure}

The $H_d^0$ exchange that induces CP violation in $B_s\!-\!\bar{B}_s$ mixing
contributes to the $B_s\!\!\to\!\mu^+\mu^-$ branching fraction, provided the coupling of $H_d^0$ to muons is 
not negligible. The coupling $y_{\mu} H_d^0 \, \bar{\mu}_R\mu_L$ leads to \\ [-5mm]
\bear
{\cal B} \!\left( B_s\!\to\!\mu^+\mu^- \!\right) \!&=&\! (|y_{bs}|^2 + |y_{sb}|^2) \frac{|y_{\mu}|^2}{M_A^4} 
\, \frac{\eta_b^2  \tau_{B_s} M_{B_s}^5 f^2_{B_s}}{64\pi (m_b+m_s)^2}  
\nonumber \\ [1mm]
& \hspace*{-2.6cm}  \approx & \hspace*{-1.4cm} 
1.3\!\times\! 10^{-8} 
\left( \frac{|y_{bs}|}{10^{-2}} \right)^{\! 2} \! \left( \frac{|y_{\mu}|}{10^{-2}} \right)^{\! 2} \! 
\left( \frac{1\; {\rm TeV}}{M_A} \right)^{\! 4} \! .
\eear
QCD corrections are taken into account by $\eta_b \approx 1.5$ \cite{Babu:1999hn}.
The experimental limit ${\cal B}\left( B_s\!\to\! \mu^+\mu^- \right) < 4.3 \times 10^{-8}$ \cite{CDFnote9892}
imposes $|y_\mu| < 0.018$ for $M_A = 1$ TeV.
Given this constraint, the impact on $B \to K\mu^+\mu^-$ observables is relatively small \cite{Bobeth:2007dw}.

\bigskip
{\bf  Uplifted supersymmetry.}---Let us describe a renormalizable gauge-invariant 
theory that includes the interactions of Eq.~(\ref{interactions}) without violating 
current limits on flavor processes.
The MSSM parameter space contains a region where the down-type fermion 
masses are induced at one loop by the VEV of the up-type Higgs doublet $H_u$.
In this so-called uplifted Higgs region 
\cite{Dobrescu:2010mk,Altmannshofer:2010zt,Kadota:2010xm} the ratio of 
$H_u$ and $H_d$ VEVs is very large, $v_u/v_d \!\equiv\!\tan\beta\gtrsim 100$,
but all Yukawa couplings remain perturbative. 
The physical states of this uplifted two Higgs doublet model include a SM-like Higgs boson, $h^0$, 
which is entirely part of $H_u$ in the $\tan\beta\!\to\!\infty$ limit, the two neutral states $H^0$ and $A^0$ of mass $M_A$,
and a charged Higgs boson $H^\pm$ of mass $M_{H^+} = (M_A^2 + M_W^2)^{1/2} \approx M_A$. The heavy states $H^0$, $A^0$ and $H^\pm$ are almost entirely part of $H_d$.

The Yukawa terms in the superpotential give rise to $H_d$ couplings to down-type 
fermions in the Lagrangian:
\be
- H_d \left( d^c \hat{y}_d Q + e^c   \hat{y}_\ell L \right) + {\rm H.c.} ~~,
\label{down}
\ee
where the quark and leptons shown here are gauge eigenstates, 
and their generation index is implicit.
The $\hat{y}_d$ and $\hat{y}_\ell$ couplings are $3\times 3$ matrices in flavor space.
Various 1-loop diagrams involving superpartners generate couplings of $H_u^\dagger$ to down-type fermions,
\be
-  H_u^\dagger \left( d^c  \hat{y}^\prime_d Q + e^c  \hat{y}^\prime_\ell L \right) + {\rm H.c.}~,
\label{wrong}
\ee
inducing masses for down-type quarks and charged leptons. The dominant contributions, from gluino and wino loops,
to the effective quark Yukawa matrix  are
\be
(\hat{y}_d^\prime)_{ij}  \approx  -\frac{\alpha_s}{4\pi}e^{-i\theta_\mu} (\hat{y}_d)_{ij}   f_{ij}  ~~.
\ee
The complex coefficients $f_{ij}$ have magnitude of order one: 
\be
f_{ij}  \approx \frac{8  |\mu| e^{i\theta_{\tilde{g}} }\!\!}{3M_{\tilde{d_i}}} \,
F\!\left(  \!\frac{M_{\tilde{g}}}{M_{\tilde{Q}_j}},\frac{M_{\tilde{d}_i}}{M_{\tilde{Q}_j}} \!\!\right) 
- \frac{3\alpha e^{i\theta_{\tilde{W}} } \!}{2s_W^2 \alpha_s\!}
F\!\left(  \!\frac{M_{\tilde{W}}}{M_{\tilde{Q}_j}},\frac{|\mu|}{M_{\tilde{Q}_j}} \!\!\right)
\label{fij}
\ee
where $0 < F(x,y) < 1$  is a function given in Eq.~(3.2) of \cite{Dobrescu:2010mk}. The phases of the gluino and wino masses
are explicitly displayed here, so that $M_{\tilde{g}}, M_{\tilde{W}} > 0$. 

We assume that the communication of supersymmetry breaking to squarks is flavor blind.
In the absence of renormalization group (RG) effects of the Yukawa couplings, the squark  mass matrices at the weak scale
are proportional to the $3\times 3$ unit matrix, so that the $\hat{y}_d^\prime$  matrix is 
given by $\hat{y}_d$ times a complex number which depends on superpartner masses.
However, the large $t$, $b$ and $\tau$ Yukawa couplings have substantial RG effects, 
driving $M_{\tilde{Q}_3}\!\! \!< \! M_{\tilde{Q}_1}\!\! = \!\! M_{\tilde{Q}_2}\!$ and $M_{\tilde{d}_3} \!\!\!<\! M_{\tilde{d}_1}\!\! = \!\!M_{\tilde{d}_2}$,
which breaks the alignment between $\hat{y}^\prime_d$ and $\hat{y}_d$ in the $3j$ and $j3$ elements.
After diagonalization of the down-type quark masses ({\it i.e.}, of $\hat{y}_d^\prime$), the neutral component of $H_d$ 
acquires off-diagonal couplings as in  Eq.~(\ref{interactions}). 
Assuming that the unitary matrix
which transforms between the gauge and mass eigenstate bases of right-handed down-type  quarks is approximately the unit matrix 
we find
\bear
&& y_{bs} = y_0 \, (a_{33} - a_{31} ) (V_L^d)_{33} (V_L^d)_{23}^* ~~,
\nonumber \\ [1mm]
&& y_{sb} = y_0 \, \frac{m_s}{m_b} a_{13}  (V_L^d)_{23} (V_L^d)_{33}^* ~~,
\nonumber \\ 
&& y_b = y_0 \, \left[1+ a_{31} + \left(a_{33} - a_{31} \right) \left| (V_L^d)_{33}\right|^2 \right] ~~,
\eear
where $a_{ij} \equiv f_{11}/f_{ij} - 1$, and  $y_0 \equiv -e^{i\theta_\mu} 4\pi m_b/(\alpha_s v_h f_{11})$, with
$v_h \approx 174$ GeV.  The unitary matrix $V_L^d$ transforms the $d_{Li}$ quarks from gauge to mass eigenstates.

For $y_b= O(1)$ and $V_L^d \simeq (V_{\rm CKM})^\dagger $, we obtain $|y_{bs}| \approx 10^{-2}$, $|y_{sb}| = O( y_{bs}m_s/m_b)$, 
confirming the bounds used after Eq.~(\ref{ma-arg}). The combination of couplings that control $K-\bar{K}$ and $B_d - \bar{B}_d$ mixing,
\bear
&& |y_{sd}y_{ds}| = |y_{bs} y_{sb}|   \frac{m_d |V_{td}^2 \, a_{13}|}{m_b\, |a_{33}- a_{31}|} \lesssim  O(10^{-13}) ~~,
\nonumber\\
&& |y_{bd}y_{db}| = |y_{bs}y_{sb}|  \frac{m_d|V_{td}|^2}{m_s|V_{ts}|^2} \lesssim 2 \times 10^{-9} ~~,
\eear
are small enough to satisfy the limits from $\varepsilon_K$ and $a_{sl}^d$ for $M_A > 100$ GeV. 

In the uplifted Higgs region the $\tau$ Yukawa coupling to $H_d$ (at the weak scale) must  be large, 
$|y_\tau| \approx 1.3$, in order for the observed $m_\tau$ to be generated by wino and bino diagrams \cite{Dobrescu:2010mk}. 
The $b$ Yukawa coupling to $H_d$ may be smaller, $|y_b| \approx 0.5 - 1$, due to the large 
contribution to $m_b$ from a 1-loop gluino diagram. However, if there is a partial cancellation between the two terms in Eq.~(\ref{fij}),
then a larger Yukawa coupling $|y_b| > 1$ is needed.

The small $m_\mu$ leaves more room for its possible origin, and consequently
a wider range of values for $y_\mu$. If $m_\mu$ is generated entirely by
the Yukawa coupling to $H_d$, then 
$|y_\mu| \approx |y_\tau| m_\mu/m_\tau \approx 0.08$, which is compatible with the current limit on 
${\cal B}( B_s\!\to\!\mu^+\mu^- \!)$ provided $M_A \gtrsim 1.7$ TeV. 
Such a large mass would imply $\phi_s \approx 0.1$, which is too small to accommodate a significant
charge asymmetry.
On the other hand, $m_\mu$ may be due to loop-induced couplings of the muon to $H_u^\dagger$ 
which exist even for $y_\mu\to 0$. For example, in models of gauge mediate supersymmetry breaking 
\cite{Dine:1994vc}, which fit well the requirements of uplifted supersymmetry, there is a vectorlike chiral superfield $d_m$ with 
the quantum numbers of weak-singlet down-type squarks. The scalar components of this messenger superfield,
$\tilde{d}_m$ and $\tilde{d}_m^c$  may couple to the SM fermions \cite{Dine:1996xk}:  
$\kappa\, \tilde{d}_m \bar{\mu}_L^c t_L$ and $\kappa^\prime\, \tilde{d}_m^c \bar{t}_R^c \mu_R$ which at 1-loop give
\cite{Dobrescu:2008sz}
\be
m_\mu \simeq m_t \, \frac{3\kappa\kappa^\prime}{32\pi^2}  \frac{\Delta M_{\tilde{d}_m}^2}{M_{d_m}^2} ~~.
\ee
A typical splitting between the messenger scalar squared-masses is $\Delta M_{\tilde{d}_m}^2 \!\!\approx 0.2 M_{d_m}^2$, 
where $M_{d_m} \!\!\sim\! O(100)$ TeV is the messenger fermion mass.
The muon mass may be generated entirely through this mechanism if  $\kappa \kappa^\prime \approx 0.3$.
A similar mechanism is used in \cite{Nandi:2008zw}.
Thus, the $y_\mu$ coupling, which determines the heavy Higgs contribution to ${\cal B}( B_s\!\to\!\mu^+\mu^- \!)$, is sensitive 
to physics at the 100 TeV scale, and can be significantly smaller than 0.08.

The dominant contributions to $(g-2)_\mu$, due to wino-slepton diagrams,  
tend in the uplifted region to enhance the discrepancy between the SM and experiment \cite{Altmannshofer:2010zt}.
We point out, though, that the wino-slepton diagrams become small if the slepton doublet of the second generation
is sufficiently heavier than $M_{\tilde{W}}$, while the bino-slepton diagrams can explain the 
discrepancy if $y_\mu \gtrsim 10^{-2}$.

Flavor-changing charged currents due to $H^\pm$ exchange are important independent of RG effects.
The couplings 
\be
\hspace*{-2mm} \frac{m_b V_{ub} y_b }{y_b v_d + y_b^\prime v_u } H^- \bar{b}_R u_L
 +  \frac{m_\tau \, y_\tau }{y_\tau v_d + y_\tau^\prime v_u } \, H^-  \bar{\tau}_R \nu_L + {\rm H.c.} \, ,
\ee
($y_b^\prime$ and $y_\tau^\prime$ are the 33 
eigenvalues of  $\hat{y}^\prime_d$ and  $\hat{y}^\prime_e$) may significantly affect the 
rate for the $B^\pm\! \to \tau^\pm\nu$ decay:
\be
\frac{{\cal B}(B^-\!\to \tau\nu)}{{\cal B}(B^-\!\to \tau\nu)_{\rm SM} } = \left|1 - y_b^* y_\tau \frac{ v_h^2}{m_b m_\tau} 
\frac{M_{B^+}^2}{M_{H^+}^2}\right|^2  ~~.
\ee
Unlike the usual MSSM where  ${\cal B}(B^-\!\!\to\tau\nu)$ is smaller than in the SM, 
the uplifted region allows an enhancement compared to the SM \cite{Altmannshofer:2010zt, Pich:2009sp}, 
depending on the phase of $y_b^* y_\tau$.
This is interesting because the measurement of this branching fraction is  larger than the SM prediction 
by a factor of 2, a $\sim$$2 \sigma$ discrepancy \cite{Bona:2009cj}.
For $y_b^* y_\tau = -1$ and $M_{H^+}$ = 1 TeV, ${\cal B}(B^-\!\!\to \tau\nu)$ increases by 24\% compared to the SM 
prediction.

\smallskip\smallskip
{\bf Conclusions.}---We have shown that the evidence for CP violation reported by the D0 Collaboration
may be explained in part by the exchange of the neutral states of a two Higgs doublet model 
contributing to  $B_s\!-\!\bar{B}_s$ mixing. 
In particular, in the uplifted Higgs region of the MSSM \cite{Dobrescu:2010mk}, a large CP-violating effect in $B_s\!-\!\bar{B}_s$ mixing implies
that the $B_s\!\to\! \mu^+\mu^-$ decay could be discovered in the near future, and that, unlike in the usual MSSM, the rate for 
$B^-\!\!\to\! \tau\nu$ may be enhanced compared to the SM prediction.
Independent of the new physics interpretation, however,
the reported central value of the charge asymmetry requires 
new physics beyond $B_s\!-\!\bar{B}_s$ mixing,
for example in $\Delta B =1 $ transitions or in $B_d\!-\!\bar{B}_d$ mixing.

\smallskip\medskip
{\it Acknowledgments:} We thank C.~Bauer, L.~Dixon, E.~Eichten, E.~Gamiz, E.~Lunghi, and A.~Petrov for helpful discussions.
We are grateful to A. Buras and A.~Kronfeld for comments on the manuscript.
Fermilab is operated by Fermi Research Alliance, LLC, under 
Contract DEAC02-07CH11359 with the US Department of Energy. 


\end{document}